\documentclass[reqno,a4paper,11pt]{article}
\usepackage{amsfonts,amssymb,amsmath,epsfig,graphicx}
\usepackage{mathrsfs}
\usepackage{theorem,calc} \usepackage{url} \usepackage{enumerate}
\usepackage{vmargin}
\setmarginsrb{3cm}{3cm}{3cm}{3cm}{0.2cm}{0.2cm}{0.2cm}{0.5cm}

\usepackage{tikz} \usetikzlibrary{arrows}
\usetikzlibrary{positioning,arrows,chains,matrix,scopes,fit,decorations.markings,decorations}
\tikzstyle{block}=[draw opacity=0.7,line width=1.4cm]

\newtheorem{definition}{Definition}[section]

\newtheorem{proposition}[definition]{Proposition}

\newtheorem{theorem}[definition]{Theorem}

 \numberwithin{equation}{section}

\newtheorem{rmk}{Remark}[section]

\newcommand{\beq}{\begin{equation}} \newcommand{\eeq}{\end{equation}}
\newcommand{\bea}{\begin{eqnarray}} \newcommand{\eea}{\end{eqnarray}}
\newcommand{\beano}{\begin{eqnarray*}}
  \newcommand{\eeano}{\end{eqnarray*}}
\newcommand{\bma}{\begin{pmatrix}} \newcommand{\ema}{\end{pmatrix}}

 \newcommand{\ZZ}{{\mathbb Z}}

\newcommand{\prf}{\underline{Proof:}\ } \newcommand{\finprf}{\null
  \hfill {\rule{5pt}{5pt}}} \newcommand{\ie}{{\it i.e.}\ }
\newcommand{\eg}{{\it e.g.}\ }

\makeatletter \def\@xfootnote[#1]{%
  \protected@xdef\@thefnmark{#1}%
  \@footnotemark\@footnotetext} \makeatother

\title{Discrete Crum's Theorems and Integrable Lattice Equations}
\date{\empty}
\author{   Cheng Zhang$^{1}$\footnote{Corresponding author, email: ch.zhang.maths@gmail.com}\,,~~~Linyu Peng$^{2,3}$\,, ~~~Da-jun Zhang$^{1}$\\ \\
  \small  ch.zhang.maths@gmail.com  ~ l.peng@aoni.waseda.jp ~ djzhang@staff.shu.edu.cn  \\ \\
  \sc $^{1}$\small Department of Mathematics \\
  \sc  \small Shanghai University\\
  \sc  \small Shanghai, 200444, China \\ \\
  \sc $^{2}$\small Waseda Institute for Advanced Study\\ 
  \sc \small Waseda University \\
  \sc \small Tokyo 169-8050, Japan\\ \\ 
  \sc $^{3}$\small Department of Applied Mechanics and Aerospace Engineering\\
  \sc  \small Waseda University\\
  \sc \small Tokyo 169-8555, Japan \\ \\
 }

\begin{document}
\maketitle

\begin{abstract}
    In this paper, we develop discrete versions of  Darboux transformations and Crum's theorems for two second order difference equations. The difference equations are discretised versions (using Darboux transformations) of the spectral problems of the KdV equation, and of the modified KdV equation or sine-Gordon equation. Considering the discrete dynamics created by Darboux transformations for the difference equations, one obtains the lattice potential KdV equation, the lattice potential modified KdV equation and the lattice Schwarzian KdV equation, that are prototypes of integrable lattice equations. It turns out that, along the discretisation processes using Darboux transformations, two families of integrable systems (the KdV family, and the modified KdV or sine-Gordon family), including their continuous, semi-discrete and lattice versions, are explicitly constructed. As direct applications of the discrete Crum's theorems, multi-soliton solutions of the lattice equations are obtained.
\vspace{.2cm}


\vspace{.2cm}

\end{abstract}
\clearpage

\section{Introduction}
The original form of  Darboux transformation appeared in Darboux's classic work  \cite{Darboux, Darb1}, when he studied the following eigenvalue problem (known as the Sturm-Liouville equation, or  the one-dimensional stationary Schr\"odinger equation), 
\begin{equation}
  \label{eq:so1}
{\cal L}\,\phi= \left(  -\partial_x^2 + u \right) \phi = \lambda \,\phi\,.
\end{equation}
The  second order   differential operator   ${\cal L}$ is  called  Schr\"odinger operator, and   $\lambda$ acts as a  spectral parameter.  Darboux stated that, under certain  transformation of the functions $\phi$ and $u$,  the form of the  equation was preserved. This characterises  Darboux transformation for the  Schr\"odinger equation, and this operation can be repeated an arbitrary number of times. A remarkable feature of Darboux's results  is that the action  of an $N$-step  Darboux transformation, \ie applying Darboux transformations to the Schr\"odinger equation $N$ times,  can be encoded into compact expressions. This  is known as the Crum's theorem \cite{Crum}, and has direct applications in quantum mechanics. 

Darboux transformation and the associated Crum's theorem  have played a pivot role in the development of integrable systems. First, they are  powerful tools for constructing multi-soliton solutions of a wide range of integrable partial differential equations (see \eg the monograph \cite{Darb2}). For instance, in the Lax formulation of the  Korteweg-de Vries (KdV) equation \cite{Lax1}, in which the Schr\"odinger equation is the spectral problem, an $N$-step Darboux transformation gives rise to an $N$-soliton solution of the KdV equation.  Repeated Darboux transformations add {\em discrete dynamics} to differential systems. As illustrated in the above example, an effect of an $N$-step Darboux transformation to the Schr\"odinger equation is to introduce a discrete variable $n$, $n = 1\,,2\,,\dots\,,N$, to the functions so that they become dependent of $n$.
Having the eigenfunction $\phi$ depending on  $x$ and  $n$, the compatibility  allows us to obtain the dressing chain equation \cite{WE, VS1}, which is an integrable differential-difference (semi-discrete)  equation. 
This idea of using Darboux transformations to derive integrable semi-discrete equations has been known in the literature \cite{Levi,Levi2}.
The discrete dynamics of the eigenfunction  $\phi$ created by repeated Darboux transformations can be casted into certain difference equations (by eliminating the continuous variable) which is the discrete analogue of its continuous counterpart.  Thus, Darboux transformation is known as an effective method to {\em discretise} the original differential system. This process was described by Shabat as  ``exact" or ``integrable discretisation" \cite{Shabat1}.

Darboux transformation has also found many  applications in various aspects in mathematics: it has a nice geometric interpretation originated from the classical theory of surfaces \cite{Darb1} (see also \cite{Roger, Gu}); it initiated researches in bispectral problems of differential equations \cite{Duis1}; it can  be used to   generate other types of  explicit solutions of integrable equations such as rational solutions \cite{Darb2} and finite-gap-type solutions   \cite{VS1} (see also the monograph \cite{Dok}), {\em etc}.

In this paper, we develop Darboux transformations for certain second order difference equations following the factorisation method initially proposed by Darboux \cite{Darboux, Darb1} (other applications can be found in \cite{Fordy, Fordy2}). The first equation we consider is the discrete Schr\"odinger equation, obtained using exact discretisation of Eq \eqref{eq:so1}. In particular, we are able to express the effect of $N$-step Darboux transformations in compact forms. This result is reminiscent of the  Crum's theorem  for the Schr\"odinger equation, and  will be referred to as the {\em discrete  Crum's theorem}. 
The second difference equation is obtained using exact discretisation of another second order differential equation \eqref{eq:lilap}, which is the spectral problem of the modified KdV equation and of the sine-Gordon equation. Darboux transformation for this difference equation and the associated discrete Crum's theorem  are also  obtained.  
Moreover, in viewing Darboux transformation as {\em new} discrete dynamics to the original difference system, we can obtain some difference-difference systems using the compatibility of the two discrete variables.

The results we are presenting here have a very natural connection to integrable discrete (lattice) equations. Research in integrable discrete equations has enjoyed rapid developments over the past twenty years (see \eg \cite{BS22, HNJ1}). Important  achievements, such as the introduction of three-dimensional consistency property \cite{Nijh2, ABS3}, the classification of integrable lattice equations \cite{ABS1, ABS2} (known as the Adler-Bobenko-Suris classification), the derivation of multi-soliton solutions \cite{Nijh3, HZ}, {\em etc.}, have been established. 

The difference-difference systems derived using Darboux transformations are closely connected to certain integrable lattice equations.
Precisely, the first difference-difference system derived from the discrete Schr\"odinger equation is  connected to the lattice potential KdV (lpKdV) equation; the second system is connected to both the lattice potential modified KdV (lpmKdV) equation and the lattice Schwarzian KdV (lSKdV) equation (also called the  cross-ratio equation). 
These equations are prototypes of lattice integrable equations of the KdV type (see the review paper \cite{NC2} for their origins and properties), and  appeared in the Adler-Bobenko-Suris classification (named H1, H3$^{\delta=0}$ and  Q1$^{\delta=0}$ respectively). 
It turns out, somehow unsurprisingly, that two families of integrable equations including their continuous, semi-discrete and lattice versions are explicitly constructed by the exact discretisaion processes using Darboux transformations.  Note that examples of using Darboux transformations to derive integrable lattice equations already exist (see \eg \cite{CZ}). As interesting by-products, the three-dimensional consistency properties of these lattice equations are well accommodated into the Darboux scheme, and their soliton solutions can be obtained thanks to the discrete Crum's theorems. 

It is worth noting that Darboux transformations and the associated Crum's theorems for difference spectral problems have been investigated since the early development of integrable systems: the generic case was constructed in the pioneering paper \cite{Mat1} (although explicit formulae of the potential functions were not shown\footnote{On  page $219$, the author wrote ``\em{We do not present here the similar but more complicated formulae for all $\varphi_{[m],k}(n,t)$}" .}), and special reductions, mainly in the context of solving semi-discrete integrable equations, can be found for instance in the monograph \cite{Darb2} and references therein. However, our approach is along the exact discretisation processes, and the so-constructed lattice equations coincide with three-dimensionally consistent equations. On the one hand, this reveals the inherent connections among continuous, semi-discrete and lattice equations from the same integrable family; on the other hand,  soliton solutions of those integrable lattice equations can be directly obtained as applications of the Crum's theorems.

The paper is organised as follows: in Section $2$, we derive Darboux transformation and discrete Crum's theorem for the discrete Schr\"odinger equation; study of the lpKdV equation including its derivations and soliton solutions is presented in Section $3$; in Section $4$,  Darboux transformation and discrete Crum's theorem for another second order difference equation is constructed; these results lay the basis for studies of the lpmKdV and lSKdV equations, which are presented in Section $5$.

\section{The discrete Schr\"odinger equation and  discrete Crum's theorem}
In this section, we  consider the following second order difference equation
\begin{equation}
\label{eq:dso1}
  L\, \phi = (-T^2 - h \,T+a)\, \phi = \lambda \,\phi\,.
\end{equation}
This equation is the discrete Schr\"odinger equation \cite{Shabat1} ($L$ is also called Shabat operator) for it is obtained through exact discretisation of its continuous counterpart \eqref{eq:so1}.  
Here,  $T$ is the shift operator in $n$, $n\in \ZZ$,   $\phi$ and $h$ are   functions of $n$, and $a$ is a parameter also depending on $n$. For simplicity, we use the ~$\widetilde{}$~ notation to denote shifts in $n$. Note that, a discrete inverse scattering transform for this equation was developed \cite{BPPS, BJ}.
\subsection{The discrete Schr\"odinger equation}
We  briefly recall the derivation of the discrete Schr\"odinger equation \eqref{eq:dso1}. Details can be found in \cite{Shabat1}. Darboux's derivation of the Darboux transformation lies on the following  decomposition of  the differential operator ${\cal L}$ given by \eqref{eq:so1}
\begin{equation}\label{eq:dec1}
  {\cal L} =   -\partial_x^2 + u  = -(\partial_x + v)\,(\partial_x - v) + a \,,
\end{equation}
where $v = (\log\psi)_x= \psi_x\,\psi^{-1}$
with $\psi$ being a fixed solution of  \eqref{eq:so1} at $\lambda = a$. This decomposition is subject to
\begin{equation}
\label{eq:ric1}
  v_x+v^2+a=u\,,
\end{equation}
which is a Riccati equation for $v$.  By interchanging the two factors in \eqref{eq:dec1}, one can define a new Schr\"odinger operator $\widetilde{\cal L}$ in the form
\begin{equation}\label{eq:dec2}
\widetilde{\cal L} = -\partial_x^2 + \widetilde{u} = -(\partial_x - v)\,(\partial_x + v) + a \,,
\end{equation}
which is subject to another constraint
\begin{equation}
\label{eq:ric2}
  -v_x+v^2+a=\widetilde{u}\,.
\end{equation}
Here $\widetilde{u}$ is the {\em newly} transformed function in $\widetilde{\cal L}$. Letting
\begin{equation}
  \label{eq:darbtr1}
\widetilde{\phi} =( \partial_x -v)\,\phi\,,
\end{equation}
one can immediately show  $\widetilde{\cal L} \,\widetilde{\phi} = \lambda\,\widetilde{\phi}$.  Now Eqs \eqref{eq:ric1} and \eqref{eq:ric2} result in $\widetilde{u} =  u -  2\,v_{x}
$, and together with the expression of $\widetilde{\phi}$ in \eqref{eq:darbtr1}, one obtains the map  $(\phi,u)\mapsto  ( \widetilde{\phi},\widetilde{u})$  characterising  the Darboux transformation for the Schr\"odinger equation.  It is clear that this map is a direct consequence of the decompositions \eqref{eq:dec1} and \eqref{eq:dec2}.

A two-step Darboux transformation is to repeat the above construction  with respect to the operator $\widetilde{\cal L}$. The function $\phi$ is mapped to
\begin{equation}\label{eq:dec3}
\widetilde{\widetilde{\phi}} = (\partial_x - \widetilde{v})\,\widetilde{\phi} =  (\partial_x - \widetilde{v})\,(\partial_x - v)\,\phi\,.
\end{equation}
Eliminating $\phi_x$ and $\phi_{xx}$ in the above expression using the above formulae yields
\begin{equation}
-  \widetilde{\widetilde{\phi}} - (\widetilde{v} +v)\,\widetilde{\phi} +  a\, \phi = \lambda\, \phi\,,
\end{equation}
which is Eq \eqref{eq:dso1} with $h = \widetilde{v} +v$.  An $N$-step Darboux transformation creates $N$ shifts for $\phi$ and $u$ in the ~$\widetilde{}$~ direction. Now   $u$ and $v$ are functions of variables $x$ and $n$. Eliminating $u$ in \eqref{eq:ric1} and \eqref{eq:ric2} leads to the dressing chain equation \cite{VS1}
\begin{equation}
  \left(\widetilde{v}+v\right)_x = v^2 -\widetilde{v}^2 +a -\widetilde{a}\,,
\end{equation}
which is also the auto-B\"acklund transformation of the modified KdV equation \cite{WE}.  
\subsection{Darboux transformation for  the discrete Schr\"odinger equation}
The construction of  Darboux transformation for the   discrete Schr\"odinger equation is also based on certain decompositions of the discrete Schr\"odinger operator $L$  \eqref{eq:dso1}.
Assume that  $L$  can be decomposed into the following form
\begin{equation}
\label{eq:facto}
  L = -(T+f)\,(T-g)+ b  \,.
\end{equation}
Here,  $f$ and $g$ are functions of $n$, and $b$ is a parameter independent of $n$.
For this decomposition to hold, one needs
\begin{equation}
  \label{eq:f1111}
  -f+\widetilde{ g}+h=0\,,\quad f\,g =a-b \,.
\end{equation}
Eliminating $f$ yields
\begin{equation}
  \label{eq:dricc1}
  (\widetilde{ g} + h)\,g+b-a=0\,,
\end{equation}
which is a discrete Riccati equation for $g$.  Posing $ g = {\widetilde{\psi}}\,{\psi}^{-1}$,
with $\psi$ be a fixed solution of Eq~\eqref{eq:dso1} at $\lambda  = b$ solves  Eq~\eqref{eq:dricc1}.   The one-step Darboux transformation for \eqref{eq:dso1} can be constructed by interchanging the two factors in the decomposition \eqref{eq:facto}.
\begin{theorem}\label{th:21}
  Consider the discrete Schr\"odinger equation \eqref{eq:dso1}.
  Under the following map
\begin{align}
\label{eq:darb2}
  \phi\mapsto & ~ \widehat{\phi} = \widetilde{\phi}-g \,\phi\,,\\
  \label{eq:darb22}
  h\mapsto & ~ \widehat{h} = \widetilde{h} + \widetilde{\widetilde{g}}-g\,,
\end{align}
where $  g = \widetilde{\psi}\,\psi^{-1}$, with  $\psi$ being a fixed solution of Eq~\eqref{eq:dso1} at $\lambda= b$,  the functions $\widehat{\phi}$ and $ \widehat{h}$ satisfy
\begin{equation}
\label{eq:hL1}
  \widehat{L} \, \widehat{\phi} = (-T^2 -   \widehat{h}\,T+a)\,  \widehat{\phi} = \lambda \, \widehat{\phi}\,,
  \end{equation}
    which is also a discrete  Schr\"odinger equation.
\end{theorem}
\prf Let $\widehat{L}$ be in the form
\begin{equation}
  \label{eq:facto1}
  \widehat{L} =  -(T-g)\,(T+f)+ b  \,.
\end{equation}
Taking account of the  decomposition~\eqref{eq:facto}, direct calculation shows $\widehat{L} \, \widehat{\phi}  =  \lambda \, \widehat{\phi}$, with $\widehat{\phi}$ given in \eqref{eq:darb2}. Equating the two expressions of $\widehat{L}$, namely Eqs \eqref{eq:hL1} and  \eqref{eq:facto1}, one gets
\begin{equation}
  \label{eq:g1111}
    -\widetilde{f}+{ g}+\widehat{h}=0\,,\quad f\,g =a-b \,.
  \end{equation}
  Combining these two equations with \eqref{eq:f1111} yields the map $h\mapsto \widehat{h}$ given in \eqref{eq:darb22}.
\finprf

Again the decompositions \eqref{eq:facto} and \eqref{eq:facto1} are essential to characterise Darboux transformation for the discrete  Schr\"odinger equation \eqref{eq:dso1}. We will use the $~\widehat{}~$ notation to denote the shifts created by the repeated Darboux transformations, and a discrete variable $m$, $m \in \ZZ$, is assigned to this direction. Clearly, starting from \eqref{eq:dso1}, its Darboux transformation adds another discrete variable to the system, and the functions $h$ and $\phi$ now depend on both $n$ and $m$ (and also on $x$ as they descend from a differential system).

\subsection{$N$-step Darboux transformation and discrete Crum's theorem}
As an illustration, we first  show an explicit construction of a two-step Darboux transformation. Let $\psi_1$ and $\psi_2$  be two linearly independent solutions of \eqref{eq:dso1}. One has
\begin{equation}
  \widehat{\psi}_2  = \widetilde{\psi}_2-g_1\,\psi_2\,,
\end{equation}
as a  solution of Eq \eqref{eq:hL1}
 with $g_1= \widetilde{\psi}_1\,\psi_1^{-1}$. Applying again Theorem \ref{th:21} to $\widehat{L}$ yields
\begin{align}
  \widehat{\widehat{\phi}} &=\widetilde{\widehat{\phi}}-g_2\,\widehat{\phi}=(T-g_2)\,(T-g_1)\,\phi\,,\\
  \widehat{\widehat{h}}&=\widetilde{\widehat{h}}+\widetilde{\widetilde{g}}_2 -g_2 = \widetilde{\widetilde{h}}+\widetilde{\widetilde{\widetilde{g}}}_1 -\widetilde{g}_1+\widetilde{\widetilde{g}}_2 -g_2\,,
\end{align}
where $g_2=   \widetilde{\widehat{\psi}}_2 \, \widehat{\psi}_2^{-1}$.  In particular, it is easy to show $
    \widehat{\widehat{\psi}}_j =0$, $j = 1,2$. Recursively, given $N$ linearly independent solutions $\psi_j$  of \eqref{eq:dso1}, $j=1\,,2\,,\dots\,,N$, one can construct the  $N$-step  map $(\phi, h)\mapsto (\phi[N], h[N])$. The next theorem states that $\phi[N]$ and $ h[N]$ admit compact expressions, similar to the Crum's theorem.

\begin{theorem}\label{th:22} {\bf (Discrete Crum's Theorem)} Assuming there are $N$  fixed solutions $\psi_j$  of  the discrete Schr\"odinger equation \eqref{eq:dso1} associated with $N$ distinct parameters $\lambda = b_j$, $j = 1\,,2\,,\dots\,,N$.  Then  the $N$-step Darboux transformation amounts to the following map
\begin{align}
\label{eq:PHIN0}
\phi\mapsto  \phi[N] &=  \frac{C(\psi_1,\psi_2,\dots, \psi_N, \phi) }{C(\psi_1,\psi_2,\dots, \psi_N)} \,,\\
\label{eq:HN0}
h\mapsto  h[N] &=  h^{(N)} -s_1^{(2)}+s_1 \,.
\end{align}
Here the superscript $^{(N)}$ denotes $N$  shifts in the ~$\widetilde{}$~ direction,  $C(\varphi_1,\varphi_2,\dots, \varphi_l)$ is the Casorati determinant for functions $\varphi_1\,, \varphi_2\,,\, \dots\,,\, \varphi_l$, namely
\begin{equation}
  \label{CASO}
  C(\varphi_1,\varphi_2,\dots, \varphi_l) = \begin{vmatrix} \varphi_1 & \varphi_1^{(1)} & \dots & \varphi_1^{(l-1)} \\
\vdots &\vdots & \dots & \vdots   \\
\varphi_l & \varphi_l^{(1)} & \dots & \varphi_l^{(l-1)} \end{vmatrix} \, ,
\end{equation}
and $s_1$ is in the form
\begin{equation}
  \label{eq:HN1}
  s_1= - \frac{\begin{vmatrix} \psi_1 & \psi_1^{(1)} & \dots & \psi_1^{(N-2)} & \psi_1^{(N)}  \\
\vdots &\vdots & \dots & \vdots& \vdots   \\
\psi_N & \psi_N^{(1)} & \dots & \psi_N^{(N-2)} & \psi_N^{(N)} \end{vmatrix}}{ C(\psi_1,\psi_2,\dots, \psi_{N})}\,.
\end{equation}
\end{theorem}
\prf Setting $D[N] =(T-g_N)\,(T-g_{N-1})\,\cdots\, (T-g_1)$, then  $  \phi[N] = D[N]\, \phi$. One can expand $\phi[N]$ in a general form
\begin{align}
  \phi[N] = \phi^{(N)} +s_1 \,\phi^{(N-1)}  +s_2 \,\phi^{(N-2)}+\dots +  s_N\, \phi\,,
\end{align}
with  functions  $s_1, s_2, \dots, s_N$ to be determined. Insert this expression to
\begin{equation}
{L}[N] \, {\phi}[N] = (-T^2 -   {h}[N]\,T+a)\,  {\phi}[N] = \lambda \, {\phi}[N]\,.
\end{equation}
By equating the coefficient of $\phi^{(N+1)}$,  one obtains the formula of $h[N]$  in \eqref{eq:HN0}.
It remains to find the expressions of the coefficient functions $s_j$, $j =1\,,2\,,\dots\,,N$.  This can be done  using the following set of equations
\begin{equation}
D[N] \,\phi\big\rvert_{\phi=\psi_j} = 0\,,\quad j= 1\,,2\,,\dots\,,N\,,
\end{equation}
or in  matrix form
\begin{equation}
  \bma \psi_1 & \psi_1^{(1)} & \dots & \psi_1^{(N-1)} \\
\vdots &\vdots & \dots & \vdots   \\
\psi_N & \psi_N^{(1)} & \dots & \psi_N^{(N-1)} \ema \, \bma s_N \\ \vdots \\ s_1 \ema = -\bma \psi_1^{(N)} \\ \vdots \\ \psi_N^{(N)}  \ema\,.
\end{equation}
Solving this system of equations  using the Cramer's rule  leads to  \eqref{eq:PHIN0} and \eqref{eq:HN1}.

\begin{figure}[th]
  \centering
  \begin{tikzpicture}[scale=0.6, decoration={markings,mark=at position 0.55 with {\arrow{latex}}}]
    \def\lx{4}%
    \def\ly{2}%
    \coordinate (u00) at (0,0);
    \coordinate (u10) at (\lx,\ly);
    \coordinate (u01) at (\lx,-\ly);
    \coordinate (u11) at (2*\lx,0);
    \draw[postaction={decorate}] (u00) node[left] {$ (\phi, h)$} -- node[above left] {\footnotesize$\psi_1$} (u10);
    \draw[postaction={decorate}] (u00)  -- node[below left] {\footnotesize $\psi_2$} (u01);
    \draw[postaction={decorate}] (u10)  -- node[above right] {\footnotesize$\widehat{\psi}_2$} (u11);
    \draw[postaction={decorate}] (u01)  -- node[below right] {\footnotesize$\widehat{\psi}_1$} (u11) node[right] {$  ( \widehat{\widehat{\phi}}, \widehat{\widehat{h}})$};

  \end{tikzpicture}
  \caption{Bianchi permutability property: given  $\psi_1$ and $\psi_2$, there are two ways to make a two-step Darboux transformation from  $ (\phi, h)$. The final results $( \widehat{\widehat{\phi}}, \widehat{\widehat{h}})$ are the same.} \label{fig:210}
\end{figure}
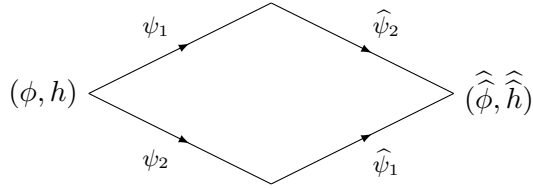

\begin{rmk}One of the most  important properties  of an $N$-step Darboux transformation is that the final result of the map  $ (\phi, h)\mapsto (\phi[N],h[N])$ is independent of the order in which the solutions $\psi_j$, $j =1\,,2\,,\dots\,,N$, are added. This property is known as the {\em  Bianchi permutability property} (see Fig.~\ref{fig:210}). This can be easily realised by looking at the determinantal forms of $\phi[N]$ and $h[N]$. Namely,  changing the positions of any two elements in the expressions  \eqref{eq:PHIN0} and \eqref{eq:HN1}, \eg $\psi_i$ and $\psi_j$, exchanges the order of adding $\psi_i$ and $\psi_j$ in the process of the $N$-step Darboux transformation, but  the final results remain unchanged.
\end{rmk}
\finprf

\section{Connection to  the lpKdV  equation }
In Section $2$, we showed that the second order difference operator $L$ is obtained through  exact discretisation of the second order differential operator $\cal L$. By developing Darboux transformation for the difference equation \eqref{eq:dso1}, one adds another discrete dynamics to the system, which is assigned with the  discrete variable $m$. This process for the function $\phi$ is depicted as follows 
\begin{center}
  \begin{tikzpicture}[scale=0.6]
    \def\l{6}%
    \def\dl{1.5}%
    \coordinate (x1) at (0,0);
    \coordinate (c1) at (\dl,0);
    \coordinate (c2) at (\dl+\l , 0);
    \coordinate (x2) at (2*\dl+\l,0);
    \coordinate (c3) at (3*\dl+\l,0);
    \coordinate (c4) at (3*\dl+2*\l,0);
    \coordinate (x3) at (4*\dl+2*\l,0);
    \node at (x1) {$  \phi(x;\lambda)$};
    \node at (x2) {$  \phi(x,n;\lambda)$};
    \node at (x3) {~~~~~~~~$  \phi(x,n,m;\lambda)\,.$};
    \draw[->] (c1) -- node[above] {{\em Darboux transf. }for  } node[below] {${\cal L} \, \phi =\lambda \,\phi$} (c2);
    \draw[->] (c3) -- node[above] {{\em Darboux transf.} for } node[below] {${L} \, \phi =\lambda \,\phi$}(c4);
  \end{tikzpicture}
\end{center}
This section aims to explore the difference-difference systems of variables $n$ and $m$. 
\subsection{Lax pair}
It it natural to consider  the compatibility condition for $\phi$ in the difference systems 
\begin{align}\label{eq:31}
-  \widetilde{\widetilde{\phi}} & - h\,\widetilde{\phi}+a\,\phi =\lambda\, \phi\,,\\ \label{eq:32} \widehat{\phi} & = \widetilde{\phi}-g\, \phi\,.
\end{align}
This amounts to  the following system of difference equations involving $g$ and $h$
\begin{align}
  \label{eq:hh1}
  \widehat{h} - \widetilde{h} &= \widetilde{\widetilde{g}} - g \,, \\
   \label{eq:hhhhh1}
   (  \widetilde{h}+ \widetilde{\widetilde{g}} )\,h  +  \widetilde{a} & = (h + \widetilde{g})\,\widehat{h} +a \,.
\end{align}
Now,   assume   $g$ and  $h$ are expressions  of certain function $w$ (also depending on $n$ and $m$) in the forms
\begin{equation}
  \label{eqq:f1}
   h = \widetilde{\widetilde{w}} - w\,, \quad g = \widehat{w} - \widetilde{w}\,.
 \end{equation}
This assumption is perfectly legitimate, thanks to the specific form of Eq \eqref{eq:hh1}. One can thus  reduce the  system \eqref{eq:hhhhh1} into a  lattice equation involving only $w$.   Recall Eq \eqref{eq:f1111} which also involves $g$ and $h$. The forms of $g$ and $h$ \eqref{eqq:f1} impose   $ f =  \widehat{\widetilde{w}} - w$, and the relation   $  f\,g =a-b$ then reads
\begin{equation}
  \label{eq:NONH1}
  (w-\widehat{\widetilde{w}})\,(\widetilde{w}-\widehat{w}) = a-b\,,
\end{equation}
which is   the  lpKdV equation.  Here  the parameters $a$ and $b $ play the role of lattice parameters, that depend {\em a priori}  on the ~$\widetilde{}$~ and ~$\widehat{}$~  directions respectively,  thus the  lpKdV equation we derived here is non-autonomous. The system of linear difference equations  \eqref{eq:31} and \eqref{eq:32} are the Lax pair of the lpKdV equation. 

One can easily formulate the Lax pair in matrix forms.   Set a vector function $\Phi^T = (\phi\,,\widetilde{\phi})$, then Eqs \eqref{eq:31} and \eqref{eq:32} lead to
  \begin{equation}
   \widetilde{\Phi} =N\,\Phi = \bma 0 & 1 \\ a-\lambda & -h \ema\,\Phi,\quad    \widehat{\Phi} = M\,\Phi = \bma    -g & 1 \\ a-\lambda &-\widetilde{g} - h  \ema\,\Phi\,.
 \end{equation}
  Their compatibility  is equivalent to  the discrete zero-curvature condition
    \begin{equation}
    \widetilde{M}\,N = \widehat{N}\,M\,.
  \end{equation}
  \begin{rmk}
In the language of B\"acklund transformations, similar derivation of the lpKdV equation exists \cite{WE}, known as  the {\em  nonlinear superposition formula} .
  \end{rmk}
  \begin{rmk}
The expressions  $f = \widehat{\widetilde{w}} -w$ and $g = \widehat{w}-\widetilde{w}$ yield the identity
\begin{equation}
  -\widetilde{f}+\widehat{f}-\widehat{\widetilde{g}}+g = 0\,.
\end{equation}
This equation, together with   $f\,g =a-b$, is  the non-autonomous discrete  KdV equation.
\end{rmk}

\begin{figure}[h]
  \centering
  \begin{tikzpicture}[scale=.7, decoration={markings,mark=at position 0.55 with {\arrow{latex}}}]
    \def\lx{3}%
    \def\ly{1.22}%
    \def\lz{ (sqrt(\x*\x+\y*\y))}%
    \def\l{4}%
    \def\d{4}%
    \coordinate (u00) at (0,0);
    \coordinate (u10) at (\l,0);
    \coordinate (u01) at (\lx,\ly);
    \coordinate (u11) at (\l+\lx,\ly);
    \coordinate (v00) at (0,\d);
    \coordinate (v10) at (\l,\d);
    \coordinate (v01) at (\lx,\d+\ly);
    \coordinate (v11) at (\l+\lx,\d+\ly);
    \draw[-] (u00) node[below] {$w$} -- node [below]{$a$} (u10);
   \draw[-]  (u10)  node[below] {$\widetilde{w}$} -- (u11) node[right, below] {$\widehat{\widetilde{w}}$};
   \draw [dashed]  (u01) node[right, below] {$\widehat{w}$} -- (u11);
   \draw [dashed]  (u00)--node [above]{$b$}(u01) ;
    \draw[-] (v00) node[above ] {$\overline{w}$}--  (v10) node[above ] {$\widetilde{\overline{w}}$} -- (v11)node[ right, above] {$\widetilde{\widehat{\overline{w}}}$} -- (v01)node[left, above] {$\widehat{\overline{w}}$} -- (v00);
    \draw[-] (u11) --  (v11);
    \draw[-] (u00) -- node [left]{$c$}(v00);
    \draw[-] (u10) -- (v10);
    \draw[dashed] (u01) -- (v01);
    \coordinate (u011) at (1.5*\lx,1.5* \ly);
    \draw[dashed] (u01) -- (u011) ;
    \draw[dashed] (u01) -- (.66*\lx,\ly) ;
    \draw[-] (u00) -- (-0.5*\lx,-0.5* \ly);
    \draw[-] (u00) -- (-.33*\lx,0);
    \draw[-] (u10) -- (0.33*\lx+\l,0);
    \draw[-] (u10) -- (-.5*\lx+\l,-.5*\ly);
    \draw[-] (u11) -- (1.33*\lx+\l,\ly);
    \draw[-] (u11) -- (-.5*\lx+\l,-.5*\ly);
    \draw[-]  (u11)-- (1.5*\lx+\l,1.5* \ly);
  \end{tikzpicture}

  \caption{Three-dimensional consistency: all six faces of the cube admit the same form of lattice equation. } \label{fig:31}
\end{figure}
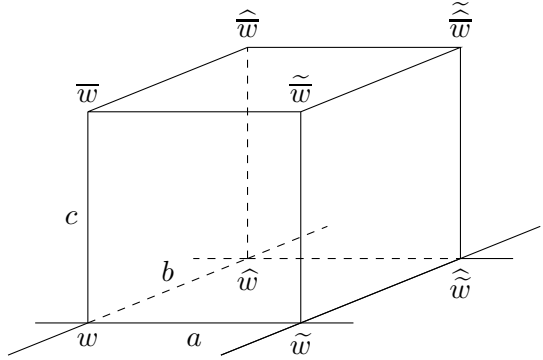

The lpKdV equation is a quadrilateral equation possessing the three-dimensional consistency property. This means  the equation can be consistently embedded onto the six faces of a cube (see Fig.~\ref{fig:31}). Here one can easily fit the  three-dimensional consistency into the Darboux derivation of the lpKdV equation.  Perform another Darboux transformation for \eqref{eq:31}, and   assign this step to a third direction, the ~$\bar{}$~  direction. Thus, one has
\begin{equation}
  \label{eq:h1111}
\overline{\phi}  = \widetilde{\phi} - {\mathcal G}  \,\phi\,, \quad  - {\mathcal F} +\overline{{\mathcal G}} +h=0\,,\quad {\mathcal F} \, {\mathcal G} = a -c\,,
\end{equation}
with
\begin{equation}
    {\mathcal F} = \widetilde{\overline{w}} -w \,,\quad {\mathcal G}  = \overline{w} -\widetilde{w}\,.
 \end{equation}
The parameter $  c$ is associated with  the  ~$\bar{}$~  direction.
Direct computations show  that the  compatibility of the two one-step Darboux transformations in the ~$\widehat{}$~   and  ~$\bar{}$~  directions  leads to the lpKdV equation
\begin{equation}
  (w-\widehat{\overline{w}})\,(\widehat{w}-\overline{w}) = b-c\,.
\end{equation}
It is clear that applying  successive Darboux transformations along three directions of the cube, the  consistency is guaranteed by the  Bianchi permutability property. 

\subsection{Soliton solutions}
Thanks to the three-dimensional consistency of the lpKdV equation, it is straightforward to apply the discrete Crum's theorem  to compute its   soliton solution. 
\begin{proposition}
  \label{prop:H11}
 Assuming there are $N$  fixed solutions $\psi_j$  associated with $N$ distinct parameters $\lambda = \lambda_j$, $j = 1\,,2\,,\dots\,,N$, of  the following system of difference equations
 \begin{align}
   \label{H1Lax1}
   -  \widetilde{\widetilde{\phi}} & - ( \widetilde{\widetilde{w}} - w)\,\widetilde{\phi}+a\,\phi =\lambda\, \phi\,,\\
   \label{H1Lax2}
   \widehat{\phi} & = \widetilde{\phi}-( \widehat{w} - \widetilde{w})\, \phi\,.
  \end{align}
  Their compatibility is the lpKdV equation \eqref{eq:NONH1}, and its $N$-soliton solution is in the form
  \begin{equation}
    \label{eq:NSH1}
w^{Nss}=w^{(N)} -s_1\,,
\end{equation}
where $s_1$ is given by Eq \eqref{eq:HN1}.

\end{proposition}
\prf
The $N$-step Darboux transformation is applied to both Eqs \eqref{H1Lax1} and  \eqref{H1Lax2} along a third direction, the ~$\bar{}$~ direction.  Using  $h = \widetilde{\widetilde{w}} - w $ and Theorem \ref{th:22}, one has
\begin{equation}
   \widetilde{\widetilde{w^{Nss}}} - w^{Nss} = w^{(N+2)} - w^{(N)} -s_1^{(2)} + s_1\,. 
\end{equation}
Solving this equation leads to the expression \eqref{eq:NSH1}.  
\finprf

For simplicity, we consider the autonomous case, \ie $a$ and $b$ are constants. A simple seed solution of the lpKdV equation is chosen as
\begin{equation}
  w = -(\alpha\,n+\beta\,m+\xi)\,,
\end{equation}
where $ \alpha^2=-a$, $ \beta^2=-b$ and $\xi$ is a constant. Then, the fixed solution  $\psi_j$ is in the form
\begin{equation}
  \psi_j=   \rho^+_j\, 
  (\alpha+\kappa_j)^n(\beta+\kappa_j)^m +\rho^-_j\,
  (\alpha-\kappa_j)^n(\beta-\kappa_j)^m \,,
\end{equation}
where  $ \kappa_j^2=-\lambda_j$ and   $\rho^\pm_j$ are  constants. The spectral parameter $\lambda_j$ acts as  the  lattice parameter in the ~$\bar{}$~ direction. Therefore, 
we recover the results  obtained  in \cite{Nijh3, HZ}.

\section{Discrete Crum's theorem: case II}
In this section, we are constructing Darboux transformation for another second order difference equation
\begin{equation}
\label{eq:2ode1}
  L\,\phi=\left(h\,T^2+(1+\widetilde{a}\,h)\,T+a\right)\,\phi = \lambda\,\phi\,.
\end{equation}
Here  $h$ is a function of $n$,  and  the parameter $a$ depends on $n$. This equation is obtained through exact discretisation from another second order differential equation \eqref{eq:lilap}. We develop one-step and  $N$-step Darboux transformations for the difference equation, and the results will  amount to another example of the  discrete Crum's theorem.  
\subsection{Derivation of Eq \eqref{eq:2ode1}}
Consider the following Schr\"odinger-type second order  differential equation
\begin{equation}\label{eq:lilap}
  {\cal L}\, \phi =   \left(  \partial_x^2 -2u_x\partial_x \right)  \phi  = \lambda \,\phi\,. 
\end{equation}
This equation is the spectral problem of  the potential modified KdV equation and the sine-Gordon equation in their Lax formulations.   Decompose $\cal L$ into the form
\begin{equation}
  {\cal L} =   \left(F\,   \partial_x+b \right)\,\left(\frac{1}{F}\,   \partial_x-a\right)+ \eta\,. 
\end{equation}
Here, the function $F$ and parameters $a$, $b$, $\eta$ are introduced to characterise the Darboux transformation. Define a new operator $\widetilde{\cal L}$ (interchanging the two factors in $\cal L$),
\begin{equation}
  {\widetilde{\cal L}} =  \left(\frac{1}{F}\,  \partial_x-a\right) (F\,   \partial_x+b )\,+ \eta\,,
\end{equation}
then clearly $  {\widetilde{\cal L}}\,\widetilde{\phi} = \lambda\,\widetilde{\phi}$, with
\begin{equation}
  \label{eq:dt2ss}
\widetilde{\phi} = \left(\frac{1}{F}\,   \partial_x-a\right) \phi\,.  
\end{equation}
Equating the two expressions of $\cal L$ and of  $ {\widetilde{\cal L}}$   (applying  a shift to $u_x$) yields
\begin{equation}
  \label{eq:FFFF}
  -\partial_x(\log F) -aF+\frac{b}{F} = -2u_x\,, \quad
  \partial_x(\log F) -aF+\frac{b}{F} = -2\widetilde{u}_x\,, \quad  a\,b=\eta\,.
\end{equation}
For simplicity, let $b =1$, thus  $\eta = a$. Arranging the above equations, one gets
\begin{equation}
  \label{eq:FFF}
  \widetilde{u}_x-u_x = -  \partial_x(\log F)\,,\quad  \widetilde{u}_x + u_x= aF-\frac{1}{F}\,.
\end{equation}
The first expression together with Eq \eqref{eq:dt2ss} define the map  $(\phi,u_x)\mapsto  ( \widetilde{\phi},\widetilde{u}_x)$, \ie Darboux transformation for Eq \eqref{eq:lilap}. Solving Eq \eqref{eq:dt2ss} and choosing  $F$ of the    particular  form
\begin{equation}
  \label{eq:semidis1}
 F=\frac{e^{(u-\widetilde{u})}}{\sqrt{a}}\,,
\end{equation}
  the second expression  of \eqref{eq:FFF} becomes
  \begin{equation}
 \left(   \frac{ \widetilde{u}+u}{2}\right)_x =\sqrt{a}\,   \sinh (u-\widetilde{u})\,,
\end{equation}
which is the semi-discrete version of  the  sinh-Gordon equation\footnote{Note that choosing $F=\frac{e^{i(u-\widetilde{u})}}{\sqrt{a}}$, one obtains the  semi-discrete  sine-Gordon equation \[ \left(   \frac{ \widetilde{u}+u}{2}\right)_x =\sqrt{-a}\,   \sin (u-\widetilde{u})\,. \]} \cite{Hirota}.

Now having obtained the one-step Darboux transformation for Eq \eqref{eq:lilap}, one can construct the  two-step Darboux transformation. A difference system  involving $\phi$, $\widetilde{\phi}$ and $\widetilde{\widetilde{\phi}}$ can be easily derived using the above formulae
\begin{equation}
  \left(F \widetilde{F} \right)\widetilde{\widetilde{\phi}}+\left(1 + \widetilde{a}F \widetilde{F} \right)\widetilde{\phi}+a\,\phi = \lambda\,\phi\,.
\end{equation}
Setting $h = F\widetilde{F}$, one recovers Eq \eqref{eq:2ode1}.

\subsection{One-step Darboux transformation}
To construct  Darboux transformation for Eq \eqref{eq:2ode1}, one  decomposes the operator $L$ into
\begin{equation}
L=(A\,T+f+c)(B\,T-g-b)+\sigma\,,
\end{equation}
where $A$, $B$, $f$ and $g$ are functions of both the variables $n$ and  $m$, corresponding to the  ~$\widetilde{}~$  and ~$\widehat{}~$   directions respectively, and $b$, $c$ and $\sigma$ are parameters independent of the  ~$\widetilde{}~$   direction  but depend {\it a priori} on the  ~$\widehat{}~$   direction.  Equating the two expressions of $L$ leads to
\begin{equation}\label{eq:sys1}
A\,\widetilde{B}=h,\quad 1+\widetilde{a}\,h=(f+c)\,B-(\widetilde{g}+b)\,A\,,\quad \sigma-a=(f+c)(g+b)\,.
\end{equation}
This  allows us to define a Darboux transformation
\begin{align}
L&\mapsto \widehat{L}=(B\,T-g-b)(A\,T+f+c)+\sigma\,,\\
\phi&\mapsto\widehat{\phi}=(B\,T-g-b)\,\phi\,,
\end{align}
obviously subject to $\widehat{L}\,\widehat{\phi}=\lambda\,\widehat{\phi}$, if  $ \widehat{h}$ satisfies the following constraints
\begin{equation}\label{eq:sys2}
\widetilde{A}\,B =\widehat{h}\,,\quad 1+\widetilde{a}\,\widehat{h}=\left(\widetilde{f}+c\right)\,B-(g+b)\,A\,.
\end{equation}
At first glance, Eqs~\eqref{eq:sys1} and \eqref{eq:sys2} define a complicated system of equations involving $A$, $B$, $f$, $g$ and $h$. However, this system can be significantly simplified. 

First, eliminating $f+c$, 
one gets
\begin{equation}  \label{eq:sys3-1}
\frac{\left(1+b\,\widetilde{A}\right)+\widetilde{A}\left(\widetilde{\widetilde{g}}+\widetilde{\widetilde{a}}\,\widetilde{\widetilde{B}}\right)}{\widetilde{B}}=\frac{(1+b\,A)+g\,A+\widetilde{a}\,\widetilde{A}\,B}{B}\,.
\end{equation}
Since we are content here to find a possible solution of the system, we can choose $
  g=-a\,B
$, which leaves  Eq \eqref{eq:sys3-1} in the form
\begin{equation}
  (T-1)\,\left(\frac{1}{B}+\frac{b\,A}{B}-a\,A\right) = 0\,.
\end{equation}
This implies $\frac{1}{B}+\frac{b\,A}{B}-a\,A =\gamma$,
with $\gamma$ being a constant of summation, which may depend on the ~$\widehat{}$~ direction.  Substituting this  expression  back to Eqs \eqref{eq:sys1} and \eqref{eq:sys2}, a last equation involving the parameters is obtained,
\begin{equation}
\sigma-\gamma\, b=0\,.
\end{equation}
Without loss of generality, one absorbs  $\gamma $ into $b$ by letting $\gamma =1$, so that  $\sigma =b$.
Thus, we have obtained
\begin{equation}
\label{eq:sys5}
\quad  g=-a\,B\,,\quad A =\frac{B-1}{b-a\,B}\,,\quad f+c=\frac{b-a}{b-a\,B}\,,
 \end{equation}
  and
  \begin{equation}
  \label{eq:sysc}
    \widehat{h}\,\widetilde{\widetilde{B}} = \widetilde{h}\,B\,,
  \end{equation}
  where $B$ satisfies certain discrete Riccati equation
\begin{equation}
\label{eq:sys7}
  -(b-a\,B)\, h  +  ( B - 1)\, \widetilde{B} = 0\,.
  \end{equation}
  Letting
  \begin{equation}\label{eq:sys8}
    B = \frac{b \,\psi}{ \widetilde{\psi} +a \,\psi}\,,
  \end{equation}
  with $\psi$ being a fixed solution of the eigenvalue problem \eqref{eq:2ode1} at $\lambda =b$, one solves the  Riccati equation \eqref{eq:sys7}.  We summarise the above analysis into the following theorem, which characterises the one-step Darboux transformation for Eq \eqref{eq:2ode1}.
  \begin{theorem}\label{prop:sodt}
Consider the second order difference equation  \eqref{eq:2ode1}. Under the following map
\begin{align}
  \label{eq:DT222}
  \phi &\mapsto\widehat{\phi}=B\,\widetilde{\phi}+(a\,B-b)\,\phi\,,\\
h  &\mapsto \widehat{h}=\frac{B\,\widetilde{h}}{\widetilde{\widetilde{B}}}\,,
\end{align}
where $B$ is defined in \eqref{eq:sys8} with $\psi$ being a fixed solution of \eqref{eq:2ode1} at $\lambda = b$, the functions $\widehat{\phi}$ and $\widehat{h}$ satisfy
\begin{equation}
\widehat{L}\,\widehat{\phi}=\left(\widehat{h}\,T^2+(1+\widetilde{a}\,\widehat{h})\,T+a\right)\,\widehat{\phi} = \lambda\,\widehat{\phi}\,,
\end{equation}
which shares the same form as Eq \eqref{eq:2ode1}.
  \end{theorem}

 \subsection{$N$-step Darboux transformation and discrete Crum's theorem}
 We can repeat the above Darboux transformation an arbitrary number of times. Interestingly,   we are able to express the $N$-step mapped functions $\phi[N]$ and $h[N]$ in compact forms. These results are referred to as the discrete Crum's theorem for Eq \eqref{eq:2ode1}.

As an illustration, let us first compute $\phi[2]$ and $h[2]$ explicitly. Assume that  $\psi_1$ and $\psi_2$ are two fixed solutions of  Eq \eqref{eq:2ode1} associated with  $\lambda=b_1$ and $b_2$ respectively, with $b_1 \neq b_2$.
Following Theorem  \ref{prop:sodt},  a one-step Darboux transformation using $\psi_1 $ leads to explicit expressions
\begin{equation}\label{phi:1dt}
\phi[1]=b_1\,\frac{ \psi_1\,\widetilde{\phi}-\widetilde{\psi}_1\, \phi}{\widetilde{\psi}_1+a\psi_1}\,,\quad h[1] = \widetilde{h}\,\frac{\psi_1}{\widetilde{\psi}_1+a\psi_1}\frac{\widetilde{\widetilde{\widetilde{\psi}}}_1+\widetilde{\widetilde{a}}\,\widetilde{\widetilde{\psi}}_1}{\widetilde{\widetilde{\psi}}_1} \,,
\end{equation}
and $\phi[1]$ satisfies
\begin{equation}
h[1]\,\widetilde{\widetilde{\phi[1]}}+\left(1+\widetilde{a}\,h[1]\right)\widetilde{\phi[1]}+a\,\phi[1]=\lambda\,\phi[1]\,.
\end{equation}
Recall that $T$ denotes the shift operator along the ~$\widetilde{}~$ direction, and $C(\varphi_1,\varphi_2,\dots, \varphi_{N})$ is the Casorati determinant,   we can rewrite \eqref{phi:1dt} as
\begin{equation}
\phi[1]=b_1\,\frac{ C(\psi_1,\phi)}{\det \mathcal{M}_1}\,,\quad h[1] =\widetilde{h}\,\frac{C(\psi_1)}{\det \mathcal{M}_1} \, T^2\left(\frac{\det\mathcal{M}_1}{C(\psi_1)}\right)\,,
\end{equation}
where the matrix $\mathcal{M}_1$ is in the form
  \begin{equation}
    {\mathcal{M}_1} = \bma 1 & -a\\
    \psi_1 & \widetilde{\psi}_1
    \ema.
  \end{equation}

  A two-step Darboux transformation is constructed using both $\psi_1$ and $\psi_2$. Precisely, one has
\begin{equation}
\phi[2]=B_2\,\widetilde{\phi[1]}+\left(a\,B_2-b_2\right)\phi[1]\,,
\end{equation}
where
\begin{equation}
B_2=\frac{b_2\,\psi_2[1]}{\widetilde{\psi_2[1]}+a\,\psi_2[1]}\,,
\quad
\psi_2[1]=b_1\frac{C(\psi_1,\psi_2)}{\det\mathcal{M}_1}\,.
\end{equation}
Combining these expressions together, we obtain
\begin{equation}
\phi[2]=b_1b_2\frac{C(\psi_1,\psi_2)\,C(\widetilde{\psi}_1,\widetilde{\phi})-C(\widetilde{\psi}_1,\widetilde{\psi}_2)\,C(\psi_1,\phi)}{C(\widetilde{\psi}_1,\widetilde{\psi}_2)\, \det \mathcal{M}_1+a\, C(\psi_1,\psi_2)\, T\det \mathcal{M}_1} =b_1b_2\frac{C(\psi_1,\psi_2,\phi)}{\det \mathcal{M}_2}\,,
\end{equation}
where
  \begin{equation}
    {\mathcal{M}_2} = \bma 1 & -a & (-a)(-\widetilde{a}) \\
    \psi_1 & \widetilde{\psi}_1  &\widetilde{\widetilde{\psi}}_1  \\
    \psi_2 & \widetilde{\psi}_2 &\widetilde{\widetilde{\psi}}_2
    \ema \,.
  \end{equation}
Similarly, the expression of $h[2]$ is in the form
  \begin{equation}
  h[2]=\widetilde{\widetilde{h}}\,\frac{C(\psi_1,\psi_2)}{\det \mathcal{M}_2}\,T^2\left(\frac{\det \mathcal{M}_2}{C(\psi_1,\psi_2)}\right)\,. 
  \end{equation}
These  give us  suggestive forms  of $\phi[N]$ and $ h[N]$,  which are stated in the following discrete Crum's theorem.
\begin{theorem} \label{theoAAA}
  Assuming that there are $N$ fixed solutions $\psi_j$ associated with $N$ distinct parameters $\lambda = b_j$, $j = 1\, , 2\, , \dots\, , N$, of the second order difference equation \eqref{eq:2ode1}. Then the $N$-step Darboux transformation amounts to the following map
  \begin{align}
    \label{eq:PHIN}
    \phi\mapsto  \phi[N] &=  \left(\prod_{j=1}^{N}b_j \right)\, \frac{C(\psi_1,\psi_2,\dots, \psi_N, \phi)}{\det \mathcal{M}_N}  \,,\\
    \label{eq:HN}
    h\mapsto  h[N] &=  h^{(N)} \, \frac{C(\psi_1,\psi_2,\dots, \psi_N)}{\det \mathcal{M}_N} \, T^2 \left( \frac{\det \mathcal{M}_N}{C(\psi_1,\psi_2,\dots, \psi_N)} \right)  \,,
  \end{align}
  where the matrix $\mathcal{M}_N$ is in the form
  \begin{equation}
    \label{eq:mmn1}
    {\mathcal{M}_N} = \bma 1 & -a & \dots &\prod^N_{j=1}\left(-a^{(j-1)}\right) \\
    \psi_1 & \psi^{(1)}_1  & \dots &\psi^{(N)}_1  \\
    \psi_2 & \psi^{(1)}_2 & \dots &\psi^{(N)}_2 \\
    \vdots & \vdots & \vdots &  \vdots \\
    \psi_N & \psi^{(1)}_N & \dots &  \psi_N^{(N)}
    \ema.
  \end{equation}
\end{theorem}
\prf
{\bf Step 1.}
It is obvious that $\phi[N]$ is linear with respect to $\phi$ and its shifts, then an $N$-step Darboux transformation gives
\begin{equation}\label{eq:phiNN}
\begin{aligned}
\phi[N]&=\left(B_N\,T+a\,B_N-b_N\right)\phi[N-1] \\
&=\left(B_N\,T+a\,B_N-b_N\right)\,\cdots\,\left(B_1\,T+a\,B_1-b_1\right)\phi \\
&=S_{N,0}\,\phi^{(N)}+S_{N,1}\,\phi^{(N-1)}+\, \cdots\, +S_{N,N}\,\phi\,,
\end{aligned}
\end{equation}
with the coefficients $S_{N,0},S_{N,1},\dots,S_{N,N}$ to be determined. It is easy to show
\begin{equation}
  \label{SSN}
  S_{N,0} =B_N\,B_{N-1}^{(1)}\, \cdots\,  B_1^{(N-1)}\,.
\end{equation}
Since $\left.\phi[N]\right|_{\phi=\psi_j}=0$, $j=1\,,2\,,\ldots\,,N$, we obtain the following system involving all $\psi_j$
\begin{equation}\label{eq:sys10}
  \bma \psi_1 & \psi_1^{(1)} & \dots & \psi_1^{(N-1)} \\
\vdots &\vdots & \dots & \vdots   \\
\psi_N & \psi_N^{(1)} & \dots & \psi_N^{(N-1)} \ema \, \bma S_{N,N} \\ \vdots \\ S_{N,1}\ema = -S_{N,0}\bma \psi_1^{(N)} \\ \vdots \\ \psi_N^{(N)}  \ema \,.
\end{equation}
Using the Cramer's rule, one obtains
\begin{equation}
S_{N,N+1-j}=-S_{N,0}\,\frac{C_j}{C(\psi_1,\psi_2,\ldots,\psi_N)}\,,\quad j=1\,,2\,,\ldots\,,N,\,
\end{equation}
where
 the function  $C_j$ denotes the Casorati determinant  $C(\psi_1,\psi_2,\ldots,\psi_N)$ by replacing its $j$-th column by the vector $\left( \psi_1^{(N)} \, \psi_2^{(N)} \,\ldots\, \psi_N^{(N)}\right)^T$. Thus, one has
\begin{equation}\label{sym:phin}
\phi[N]=S_{N,0}\,\frac{C(\psi_1,\psi_2,\ldots,\psi_N,\phi)}{C(\psi_1,\psi_2,\ldots,\psi_N)}\,.
\end{equation}
On the other hand, inserting the expression \eqref{eq:phiNN} into $L[N ] \, \phi[N] = \lambda\,\phi[N]$ gives $h[N]$ in the form
\begin{equation}\label{sym:hn}
h[N]=h^{(N)}\,\frac{S_{N,0}}{\widetilde{\widetilde{ S_{N,0}}}}\,.
\end{equation}
{ \bf Step 2.} It remains to determine $S_{N,0}$ to get explicit expressions of $\phi[N]$ and $h[N]$. 
It is clear that the $N$-step Darboux transformation yields  
\begin{equation}\label{sys:bn}
B_N=\frac{b_N\,\psi_N[N-1]}{\widetilde{\psi_N[N -1]}+a\,\psi_N[N-1]}\,. 
\end{equation}
From Eq \eqref{sym:phin}, we have
\begin{equation}
\psi_N[N-1]=S_{N-1,0}\,\frac{C(\psi_1,\psi_2,\ldots,\psi_N)}{C(\psi_1,\psi_2,\ldots,\psi_{N-1})}\,.
\end{equation}
Using $B_N= \frac{S_{N,0}}{\widetilde{S_{N-1,0}}}$ (coming from \eqref{SSN}), one gets the following identity
\begin{equation}\label{sym:trs}
\begin{aligned}
S_{N,0}\,\widetilde{ S_{N-1,0}}\,\widetilde{R_{N-1}}+a\,S_{N,0}\,S_{N-1,0}\,R_{N-1}-b_N\,S_{N-1,0}\,\widetilde{S_{N-1,0}}\,R_{N-1} =0\,,
\end{aligned}
\end{equation}
where
\begin{equation}
R_{N-1}=\frac{C(\psi_1,\psi_2,\ldots,\psi_N)}{C(\psi_1,\psi_2,\ldots,\psi_{N-1})}\,.
\end{equation}
One can prove by  induction  that $S_{N,0}$ defined in the form
\begin{equation}
  \label{eq:SNOcf}
S_{N,0}=\left(\prod_{j=1}^{N}b_j \right)\,\frac{C(\psi_1,\psi_2,\ldots,\psi_N)}{\det \mathcal{M}_N}
\end{equation}
satisfies \eqref{sym:trs}.
Substituting $S_{N,0}$ back to Eqs \eqref{sym:phin} and \eqref{sym:hn} completes the proof.\finprf
 \section{Connections to the lpmKdV  and  lSKdV equations}
 In this section, we explore the difference-difference systems derived in the previous section.  Thanks to some appropriate substitutions of the functions $h$, $A$ and $B$,  we obtain  two integrable lattice equations:  the lpmKdV equation and the lSKdV equation.  Their  $N$-soliton solutions will be derived as  consequences of Theorem \ref{theoAAA}. 
 \subsection{The lpmKdV equation}
 Recall the derivation of the semi-discrete sinh-Gordon equation (see Section $4.1$). Letting $w = e^u$ and   $\sqrt{a} = \alpha$,  from the  expression  Eq~\eqref{eq:semidis1}, one has
 \begin{equation}\label{eq:subs11}
 F =\frac{1}{\alpha } \frac{w}{\widetilde{w}}\,.    
 \end{equation}
 Also knowing that $  h = F\,\widetilde{F}$,  it is straightforward to express the functions $A$ and $B$ (defined in  Section $4.2$) in the forms
 \begin{equation}\label{eq:subs12}
 A=\frac{1}{\alpha\,\beta}\frac{w}{\widehat{\widetilde{w}}}\,, \quad B = \frac{\beta}{\alpha}\frac{\widehat{w}}{ \widetilde{w}}\,,
   \end{equation}
where $\beta$ is a parameter depending on  the ~$\widehat{}$~ direction only.  Now consider the relation between $A$ and $B$ defined in Eq \eqref{eq:sys5}. Substituting \eqref{eq:subs11} and \eqref{eq:subs12} inside, and setting $a =\alpha^2$ and $b=\beta^2$, one gets
   \begin{equation}
     \label{eq:H3cf}
     \alpha \left( w\,\widehat{w} - \widetilde{w}\,\widehat{\widetilde{w}}\right) -     \beta \left( w\,\widetilde{w}- \widehat{w}\,\widehat{\widetilde{w}}\right)=0\,,
   \end{equation}
   which is the non-autonomous lpmKdV  equation.

   The three-dimensional consistency of this equation is naturally  encoded into the Bianchi permutability property of the Darboux transformation. Namely,  perform a Darboux transformation for Eq \eqref{eq:2ode1}  in a third direction, the ~$\bar{}$~  direction. Assign $\gamma$ as the lattice parameter in this direction.   By computing the  compatibility of the two one-step Darboux transformations in  the ~$\widehat{}$~   and  ~$\bar{}$~  directions, one obtains another lpmKdV  equation
   \begin{equation}
     \alpha \left( w\,\overline{w} - \widetilde{w}\,\widetilde{\overline{w}}\right) -     \gamma \left( w\,\widetilde{w}- \overline{w}\,\widetilde{\overline{w}}\right)=0\,. 
   \end{equation}
Then, the  three-dimensional consistency of the lpmKdV  equation is a mere consequence of  successive Darboux transformations in three directions of a cube. The $N$-soliton solution can be easily obtained using Theorem  \ref{theoAAA}.
   \begin{proposition}
     \label{NSH3}
 Assuming there are $N$  fixed solutions $\psi_j$  associated with $N$ distinct parameters $\lambda = \lambda_j$, $j = 1\,,2\,,\dots\,,N$, of  the following system of difference equations
 \begin{align}
   \label{eq:H3Lax1}
    \left( \frac{1}{\alpha\,\widetilde{\alpha}}\frac{w}{\widetilde{\widetilde{w}}}\right)  \widetilde{\widetilde{\phi}} & + \left(1+ \frac{\widetilde{\alpha}}{\alpha}\frac{w}{\widetilde{\widetilde{w}}} \right)\,\widetilde{\phi}+\alpha^2\,\phi =\lambda\, \phi\,,\\
   \label{eq:H3Lax2}
   \widehat{\phi}&= \frac{\beta}{\alpha}\frac{\widehat{w}}{ \widetilde{w}}\widetilde{\phi}+\left( \alpha\,\beta\,\frac{\widehat{w}}{ \widetilde{w}}-\beta^2\right)\phi\,.
  \end{align}
  Their compatibility gives rise to the lpmKdV equation \eqref{eq:H3cf}, and its $N$-soliton solution is in the form
\begin{equation}
w^{Nss}=\left(\prod_{j=1}^{N}\alpha^{(j-1)} \right)w^{(N)}\, \frac{C(\psi_1,\psi_2,\ldots,\psi_N)}{\det \mathcal{M}_N} \,,
\end{equation}
where the matrix $\mathcal{M}_N$ is defined in Eq \eqref{eq:mmn1}.
\end{proposition}
\prf The $N$-step Darboux
transformation is applied to both  Eqs \eqref{eq:H3Lax1} and  \eqref{eq:H3Lax2} along a third direction, the ~$\bar{}$~ direction. From the above derivation of the lpmKdV  equation, it is clear  that $h = w/(\alpha\, \widetilde{\alpha}\,\widetilde{\widetilde{w}})$. Then it follows from  Theorem \ref{theoAAA} that
\begin{equation}
\frac{1}{\alpha\, \widetilde{\alpha}}\frac{w^{Nss}}{\widetilde{\widetilde{w^{Nss}}}} = \frac{1}{\alpha^{(N)}\, \alpha^{(N+1)}}\frac{w^{(N)}}{w^{(N+2)}} \frac{C(\psi_1,\psi_2,\ldots,\psi_N)}{\det \mathcal{M}_N}\, T^2\left(\frac{\det \mathcal{M}_N}{C(\psi_1,\psi_2,\ldots,\psi_N)}\right)\,.
\end{equation}
Solving this equation completes the proof. Note that without loss of generality  the constant of summation is absorbed into the seed solution $w$.
\finprf

For simplicity, consider only the autonomous case, \ie the parameters  $\alpha$ and $\beta$ are both constants. A simple seed solution is 
\begin{equation}
  w= \rho\, s^n\,t^m\,,
\end{equation}
where $s$ and $t$ are related to the lattice parameters $\alpha$ and $\beta$ (see \eqref{eq:alphabeta}), and $\rho$ is a constant. The spectral parameter  $\lambda_j$ is related to the  lattice parameter $\gamma_j$  via   $\lambda_j = \gamma_j^2$. The seed solution imposes
\begin{equation}\label{eq:alphabeta}
   \alpha = \frac{2\, \theta \,s}{s^2-1}\,,\quad \beta = \frac{2\, \theta\,t}{t^2-1}\,, \quad \gamma_j = \frac{2\, \theta\,r_j}{r_j^2-1}\,,
 \end{equation}
 where $\theta$ is a constant\footnote{Here we assume $s$, $t$ and $\theta$ are real (rather than complex). More details can be found in \cite{HZ}.}. The last term is introduced to express  $\gamma_j$ in terms of another constant $r_j$, similarly to the expressions of $\alpha$ and $\beta$. 
 Solving the Lax pair, one gets the fixed solution  $\psi_j$ in the form
 \begin{equation}
       \label{final111}
  \psi_j= \rho_j^+\,E(n,m;\kappa_j)+\rho_j^-\,E(n,m;-\kappa_j)\,,
 \end{equation}
 where $\rho^\pm_j$ are constants, and the parameter $\kappa_j$ is defined in terms of of $r_j$ as
 \begin{equation}
   \kappa_j  = \theta\,\frac{1+r_j^2}{1-r_j^2}\,.
 \end{equation}
The function $E$ is a discrete exponential-type function 
  \begin{equation}
E(n,m;\pm\kappa_j) = \left(-\frac{\alpha^2}{2}(1+s^2) \pm 2 \alpha  s\,\kappa_j\right)^n\,\left(-\frac{\beta^2}{2}(1+t^2) \pm2 \beta  t\,\kappa_j\right)^m\,.
  \end{equation}
  Without loss of generality, one can reparametrise the constants $\rho^\pm_j$ as
  \begin{equation}
    \label{final112}
    \rho^\pm_j \mapsto \rho^\pm_j \left( \frac{\alpha^2}{2}(1-s^2) \pm 2 \alpha s\,\kappa_j\right)^l\,,
  \end{equation}
  where the parameter $l$ is understood as a  discrete ``variable" in the ~$\bar{}$~ direction. Now the function  $\psi_j$ depends on $n$, $m$ and $l$ (also on the parameter $\kappa_j$). The advantage of the above parametrisation of $\psi_j$ is that one can link $\widetilde{\psi}_j$ and  $\overline{\psi}_j$ via
  \begin{equation}
    \widetilde{\psi}_j = -\alpha^2 \psi_j + \overline{\psi}_j\,,
  \end{equation}
  where $\overline{\psi}_j$ add a shift in the ~$\bar{}$~ direction by mapping the variable $l$ to $l+1$.  
 Knowing that $\alpha^2 = a$, then   the expression of  $\det \mathcal{M}_N$ can be drastically simplified to 
\begin{equation}
\label{MNN}  \det \mathcal{M}_N =  D(\overline{\psi}_1,\overline{\psi}_2,\ldots,\overline{\psi}_N)\,. 
\end{equation}
The  notation $D(\varphi_1,\varphi_2,\ldots,\varphi_N)$ denotes the Casorati determinant defined in the ~$\bar{}$~ direction, \ie $D(\varphi_1,\varphi_2,\ldots,\varphi_N)$ is in the same form as $C(\varphi_1,\varphi_2,\ldots,\varphi_N)$ (see Eq \eqref{CASO}) but with  ~$\widetilde{}$~ shifts replaced by  ~$\bar{}$~ shifts.
Using  $ \Theta_N $ to denote  $D(\psi_1,\psi_2,\ldots,\psi_N)$
, one has  $\det \mathcal{M}_N = \overline{\Theta_N}$. Moreover,  one can easily obtain 
\begin{equation}
  \Theta_N \equiv D(\psi_1,\psi_2,\ldots,\psi_N) = C(\psi_1,\psi_2,\ldots,\psi_N)  \,. 
\end{equation}
Therefore, the $N$-soliton solution  of the lpmKdV  equation is  in the from (with the constant factor being absorbed into the seed solution $w$)
\begin{equation}
  \label{eq:H3solN}
  w^{Nss} = w^{(N)} \, \Theta_N\,\overline{\Theta_N}^{-1}\,.
\end{equation}
We recover the results  obtained  in \cite{Nijh3, HZ}.  
\subsection{The lSKdV equation}
 Interestingly, the lSKdV equation can also be obtained from the Darboux transformation for Eq~\eqref{eq:2ode1}. Consider another set of substitutions of the functions $h$, $A$ and $B$\footnote{These substitutions can be obtained through some dimensional analysis.}:
\begin{equation}\label{sys:eq22}
h = \frac{\widetilde{ P}}{a\,P }  \,,\quad B = \frac{P}{Q}\,,\quad A=\frac{\widetilde{Q}}{a\,P}\,,
\end{equation}
where $P$ and $Q$ are functions of $n$ and $m$. Substituting these back to  Eqs~\eqref{eq:sys1} and \eqref{eq:sys2}, one gets two difference constraints
\begin{equation}
  \widetilde{a} \,\widehat{\widetilde{P}}\,\widetilde{P}\,Q = a \,\widehat{P}\,P\,\widetilde{\widetilde{Q}}\,,
\end{equation}
and
  \begin{equation}
  \label{eq:cons11}
    Q-\widetilde{Q} = P -\widehat{P}\,.
\end{equation}
The first difference equation  can be solved using
\begin{equation}
  (T-1)(T+1)\log Q = (T-1)\log\left(a\,\widehat{P}\,P\right)\,,
\end{equation}
and one gets
\begin{equation}
\label{eq:sys9}
  a\,\widehat{P}\,P =  b\,\widetilde{Q}\,Q\,,
\end{equation}
with $b$ being a constant of  summation  independent of the ~$\widetilde{}$~ direction.
With the aid of  a new function $z$, the second difference equation \eqref{eq:cons11} can be solved using  
\begin{equation}
\label{eq:sys6}
P  =  \widetilde{z} -z\,,\quad Q = \widehat{z} - z\,.
\end{equation}
Then, it turns out that   Eq \eqref{eq:sys9} becomes
\begin{equation}
a\left(\widehat{\widetilde{z}}-\widehat{z}\right)\left(\widetilde{z}-z\right)=b\,\left(\widetilde{\widehat{z}}-\widetilde{z}\right)\left(\widehat{z}-z\right)\,,
\end{equation}
which is  the non-autonomous  lSKdV  equation.
From the   above derivations, it is clear that the lSKdV equation has a Lax pair in the form
\begin{align}
  \label{eq:laxq01}
\left(\widetilde{\widetilde{z}}   - \widetilde{z} \right)\widetilde{\widetilde{\phi}} +\left(  \widetilde{a}\,\widetilde{\widetilde{z}} -  \widetilde{a}\,\widetilde{z} + a\,\widetilde{z} - a  \,z \right)\widetilde{\phi} + a\left( a-\lambda\right)\left(\widetilde{z}- z\right)\phi&=0  \,,\\
\label{eq:laxq02}
  \left(\widetilde{z}-z\right)\widetilde{\phi}  -\left(\widehat{z}-z\right)\widehat{\phi} +\left(a\left(\widetilde{z} -z\right)-b\left( \widehat{z}-z\right)\right)\phi&=0\,.
\end{align}

The three-dimensional consistency of this equation can be understood in the same way as showed in the cases of the lpKdV equation and the lpmKdV equation. First perform another Darboux transformation  in   a third direction, the ~$\bar{}$~  direction  assigned with a lattice parameter $c$. Computing the  compatibility of the two one-step Darboux transformations in  the ~$\widehat{}$~   and  ~$\bar{}$~  directions, one  obtains another lSKdV equation
\begin{equation}
  a\left(\widetilde{\overline{z}}-\overline{z}\right)\left(\widetilde{z}-z\right)=c\,\left(\widetilde{\overline{z}}-\widetilde{z}\right)\left(\overline{z}-z\right)\,. 
\end{equation}
Then, one can conclude that  the lSKdV equation is  three-dimensionally consistent. 

We have just shown that both the lpmKdV and lSKdV equations can be  derived from the Darboux transformation for  Eq~\eqref{eq:2ode1}.  From the substitutions \eqref{eq:subs11}, \eqref{eq:subs12} and \eqref{sys:eq22},  it follows that the lpmKdV equation and the lSKdV equation are connected using a Miura-type (B\"acklund) transformation \cite{Nijh3}
\begin{equation}\label{eq:miura1}
  \widetilde{z}-z = \frac{1}{\alpha \, w\,\widetilde{w}}\,,\quad   \widehat{z}-z = \frac{1}{\beta\,w\,\widehat{w}}\,.
\end{equation}

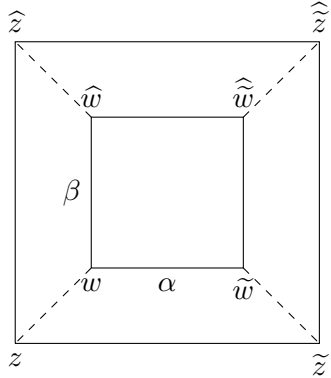
\begin{figure}[ht]
  \centering
  \begin{tikzpicture}[scale=.5, decoration={markings,mark=at position 0.55 with {\arrow{latex}}}]
 \def\l{4}%
    \def\d{2}%
       \coordinate (u00) at (0,0);
    \coordinate (u10) at (\l,0);
    \coordinate (u01) at (0,\l);
    \coordinate (u11) at (\l,\l);
    \coordinate (v00) at (-\d,-\d);
    \coordinate (v10) at (\l+\d,-\d);
    \coordinate (v01) at (-\d,\l+\d);
    \coordinate (v11) at (\l+\d,\d+\l);
    \draw[-] (u00) node[below] {$w$} -- node [below]{$\alpha$} (u10) node[below] {$\widetilde{w}$} -- (u11)  node[above] {$\widehat{\widetilde{w}}$} -- (u01) node[above] {$\widehat{w}$} -- node [left]{$\beta$}(u00);
    \draw[-] (v00) node[below] {$z$} --  (v10) node[below] {$\widetilde{z}$} -- (v11)  node[above] {$\widehat{\widetilde{z}}$} -- (v01) node[above] {$\widehat{z}$} -- (v00);
        \draw[dashed] (u00) -- (v00);
        \draw[dashed](u10) -- (v10);
        \draw[dashed](u01) -- (v01);
        \draw[dashed](u11) -- (v11);
  \end{tikzpicture}
  \caption{Miura-type transformation: consistency between two quadrilateral equations. } \label{fig:51}
\end{figure}

Now we can derive $N$-soliton solution of the lSKdV equation using that of the lpmKdV equation. Assume that $z^{Nss}$  is an $N$-soliton solution of the lSKdV equation. In addition, we assume that
$\overline{z^{Nss}}$ is a fixed-point solution of $z^{Nss}$ \cite{HZ}. Based on the three-dimensional consistency of both equations as well as the consistency between them  (see Fig. \ref{fig:51}), one can link $z^{Nss}$ and $\overline{z^{Nss}}$ to the $N$-soliton solution $w^{Nss}$ of the lpmKdV equation via similar Miura transformation
\begin{equation}
\label{fixed}
  \overline{z^{Nss}} - z^{Nss} =   \frac{ 1}{\gamma \, w^{Nss}\,\overline{w^{Nss}}}\,.
\end{equation}
Here $\gamma$  is a lattice parameter in the ~$\bar{}$~ direction that eventually depends on $a$ and $b$ in order that both the sets $(z^{Nss}, \widetilde{z^{Nss}}, \overline{z^{Nss}}, \widetilde{\overline{z^{Nss}}} )$ and $(z^{Nss}, \widehat{z^{Nss}}, \overline{z^{Nss}},\widehat{\overline{z^{Nss}}})$ satisfy the lSKdV equation. This possibly yields another relation between  $z^{Nss}$ and $\overline{z^{Nss}}$, so that one can obtain an explicit expression of $z^{Nss}$ from that of $w^{Nss}$. Since the lSKdV equation is fractional-linearly invariant, $z^{Nss}$ and $\overline{z^{Nss}}$ are simply connected by a M\"obius transformation.

For simplicity, we consider the non-autonomous case. Choosing  $\overline{z^{Nss}} = -z^{Nss}$, and taking account of the expressions \eqref{eq:H3solN} and \eqref{fixed}, one gets 
\begin{equation}
z^{Nss} = \rho \, s^{-2n} t^{-2m} \, \overline{\overline{\Theta_N}}\,\Theta_N^{-1}.
\end{equation}
Here $\rho$ is a constant (absorbing other constant factors), and $s$ and $t$ are the parameters appearing in the seed solution for  $w^{Nss}$. Note that this corresponds to the $N$-soliton solution of the lSKdV equation with  power background  obtained in \cite{Nijh3, HZ}.

\section{Concluding remarks}
By using and developing Darboux transformations for certain differential and difference equations, we manage to derive two families of integrable equations including their continuous, semi-discrete and lattice versions. These are the KdV family, including the KdV equation, the dressing chain equation and the lattice potential KdV equation; and the modified KdV or sine-Gordon family, including the modified KdV and sine-Gordon equation, the semi-discrete sinh-Gordon and sine-Gordon equation, and the lattice potential modified KdV equation and  the lattice Schwarzian KdV (cross-ratio) equation.  Darboux transformations represent in all cases an integrable discretisation process. The three-dimensional consistency properties of the lattice equations are well fitted into the Darboux scheme. The associated discrete Crum's theorems allow us to construct explicit solutions of those equations\textemdash results that are consistent with the Cauchy matrix approach \cite{Nijh3} and the bilinearisation approach \cite{HZ}.

Natural continuations of the present work are to extend our results to $1)$ other multi-dimensionally consistent equations from the Adler-Bobenko-Suris classification; $2)$ higher order differential and difference equations. Note that the decomposition (or factoristaion) methods used to derive Darboux transformations for second difference equations can also be found in general situations \cite{DJ}. It would be of interest to examine whether discrete Crum's theorems exist for those cases.

In the context of one-dimensional quantum mechanics, discrete Crum's theorem for the tridiagonal discrete Schr\"odinger equation (obtained using direct discretisation of the Schr\"odinger equation \eqref{eq:so1}) is known \cite{OS} (similar results can be found in \cite{GM1} with an integrable approach). In \cite{OS}, the authors questioned whether their version of discrete Crum's theorem have applications in discrete integrable systems. The second order difference equations (\ref{eq:dso1} and \ref{eq:2ode1})  we consider in this paper are also of Schr\"odinger-type.  Reversely, one natural question would be whether the discrete Crum's theorems we obtained in the context of discrete integrable systems can be connected to ``discrete" quantum mechanics. Note that we are also able to construct various potentials for the discrete Schr\"odinger equation \eqref{eq:dso1} that have continuous (quantum) analogues\footnote{Some results will be reported soon.}. 

After uploading this work, we received a comment from Prof.\,Liu, who, with his collaborator, have already had a different version of Crum's theorem for the discrete Schr\"odinger equation \eqref{eq:dso1} \cite{Liu} (the papar is in Chinese). This suggests that (non-trivial) Darboux transformations and the associated Crum's theorem are actually not unique.  It would be interesting to understand the connection of their results to integrable discrete systems, and also look for other possible Darboux transformations for the systems we consider in this paper.   

Lastly, it is interesting to observe that no new difference equations can be obtained using Darboux transformations for the difference equations we considered in this paper (Eqs. (\ref{eq:dso1}) and (\ref{eq:2ode1})). This is because the difference systems created by Darboux transformations remain the same as the original difference equations\textemdash a fact that is believed to be related to the discrete integrability and the multi-dimensional consistency property of the models.  

\section*{Acknowledgements}
 C.\,Zhang is supported by NSFC (No.\,11601312) and Shanghai Young Eastern Scholar program (2016-2019). L.\,Peng is partially supported by JSPS Grant-in-Aid for Scientific Research (16KT0024) and Waseda University Special Research Project (2017K-170). D.J. Zhang is supported by NSFC (No.\,11371241, 11631007). The authors would like to thank Prof.\,F.W. Nijhoff  for private communications.

\end{document}